\documentclass[onecolumn]{aastex631}
\pdfoutput=1

\emergencystretch=\maxdimen
\usepackage{lineno}
\usepackage[normalem]{ulem}

\begin{document}
\sloppy

\title{All-Sky Search for Transient Astrophysical Neutrino Emission with 10 Years of IceCube Cascade Events}

\affiliation{III. Physikalisches Institut, RWTH Aachen University, D-52056 Aachen, Germany}
\affiliation{Department of Physics, University of Adelaide, Adelaide, 5005, Australia}
\affiliation{Dept. of Physics and Astronomy, University of Alaska Anchorage, 3211 Providence Dr., Anchorage, AK 99508, USA}
\affiliation{Dept. of Physics, University of Texas at Arlington, 502 Yates St., Science Hall Rm 108, Box 19059, Arlington, TX 76019, USA}
\affiliation{CTSPS, Clark-Atlanta University, Atlanta, GA 30314, USA}
\affiliation{School of Physics and Center for Relativistic Astrophysics, Georgia Institute of Technology, Atlanta, GA 30332, USA}
\affiliation{Dept. of Physics, Southern University, Baton Rouge, LA 70813, USA}
\affiliation{Dept. of Physics, University of California, Berkeley, CA 94720, USA}
\affiliation{Lawrence Berkeley National Laboratory, Berkeley, CA 94720, USA}
\affiliation{Institut f{\"u}r Physik, Humboldt-Universit{\"a}t zu Berlin, D-12489 Berlin, Germany}
\affiliation{Fakult{\"a}t f{\"u}r Physik {\&} Astronomie, Ruhr-Universit{\"a}t Bochum, D-44780 Bochum, Germany}
\affiliation{Universit{\'e} Libre de Bruxelles, Science Faculty CP230, B-1050 Brussels, Belgium}
\affiliation{Vrije Universiteit Brussel (VUB), Dienst ELEM, B-1050 Brussels, Belgium}
\affiliation{Department of Physics and Laboratory for Particle Physics and Cosmology, Harvard University, Cambridge, MA 02138, USA}
\affiliation{Dept. of Physics, Massachusetts Institute of Technology, Cambridge, MA 02139, USA}
\affiliation{Dept. of Physics and The International Center for Hadron Astrophysics, Chiba University, Chiba 263-8522, Japan}
\affiliation{Department of Physics, Loyola University Chicago, Chicago, IL 60660, USA}
\affiliation{Dept. of Physics and Astronomy, University of Canterbury, Private Bag 4800, Christchurch, New Zealand}
\affiliation{Dept. of Physics, University of Maryland, College Park, MD 20742, USA}
\affiliation{Dept. of Astronomy, Ohio State University, Columbus, OH 43210, USA}
\affiliation{Dept. of Physics and Center for Cosmology and Astro-Particle Physics, Ohio State University, Columbus, OH 43210, USA}
\affiliation{Niels Bohr Institute, University of Copenhagen, DK-2100 Copenhagen, Denmark}
\affiliation{Dept. of Physics, TU Dortmund University, D-44221 Dortmund, Germany}
\affiliation{Dept. of Physics and Astronomy, Michigan State University, East Lansing, MI 48824, USA}
\affiliation{Dept. of Physics, University of Alberta, Edmonton, Alberta, T6G 2E1, Canada}
\affiliation{Erlangen Centre for Astroparticle Physics, Friedrich-Alexander-Universit{\"a}t Erlangen-N{\"u}rnberg, D-91058 Erlangen, Germany}
\affiliation{Physik-department, Technische Universit{\"a}t M{\"u}nchen, D-85748 Garching, Germany}
\affiliation{D{\'e}partement de physique nucl{\'e}aire et corpusculaire, Universit{\'e} de Gen{\`e}ve, CH-1211 Gen{\`e}ve, Switzerland}
\affiliation{Dept. of Physics and Astronomy, University of Gent, B-9000 Gent, Belgium}
\affiliation{Dept. of Physics and Astronomy, University of California, Irvine, CA 92697, USA}
\affiliation{Karlsruhe Institute of Technology, Institute for Astroparticle Physics, D-76021 Karlsruhe, Germany}
\affiliation{Karlsruhe Institute of Technology, Institute of Experimental Particle Physics, D-76021 Karlsruhe, Germany}
\affiliation{Dept. of Physics, Engineering Physics, and Astronomy, Queen's University, Kingston, ON K7L 3N6, Canada}
\affiliation{Department of Physics {\&} Astronomy, University of Nevada, Las Vegas, NV 89154, USA}
\affiliation{Nevada Center for Astrophysics, University of Nevada, Las Vegas, NV 89154, USA}
\affiliation{Dept. of Physics and Astronomy, University of Kansas, Lawrence, KS 66045, USA}
\affiliation{Centre for Cosmology, Particle Physics and Phenomenology - CP3, Universit{\'e} catholique de Louvain, Louvain-la-Neuve, Belgium}
\affiliation{Department of Physics, Mercer University, Macon, GA 31207-0001, USA}
\affiliation{Dept. of Astronomy, University of Wisconsin{\textemdash}Madison, Madison, WI 53706, USA}
\affiliation{Dept. of Physics and Wisconsin IceCube Particle Astrophysics Center, University of Wisconsin{\textemdash}Madison, Madison, WI 53706, USA}
\affiliation{Institute of Physics, University of Mainz, Staudinger Weg 7, D-55099 Mainz, Germany}
\affiliation{Department of Physics, Marquette University, Milwaukee, WI 53201, USA}
\affiliation{Institut f{\"u}r Kernphysik, Westf{\"a}lische Wilhelms-Universit{\"a}t M{\"u}nster, D-48149 M{\"u}nster, Germany}
\affiliation{Bartol Research Institute and Dept. of Physics and Astronomy, University of Delaware, Newark, DE 19716, USA}
\affiliation{Dept. of Physics, Yale University, New Haven, CT 06520, USA}
\affiliation{Columbia Astrophysics and Nevis Laboratories, Columbia University, New York, NY 10027, USA}
\affiliation{Dept. of Physics, University of Oxford, Parks Road, Oxford OX1 3PU, United Kingdom}
\affiliation{Dipartimento di Fisica e Astronomia Galileo Galilei, Universit{\`a} Degli Studi di Padova, I-35122 Padova PD, Italy}
\affiliation{Dept. of Physics, Drexel University, 3141 Chestnut Street, Philadelphia, PA 19104, USA}
\affiliation{Physics Department, South Dakota School of Mines and Technology, Rapid City, SD 57701, USA}
\affiliation{Dept. of Physics, University of Wisconsin, River Falls, WI 54022, USA}
\affiliation{Dept. of Physics and Astronomy, University of Rochester, Rochester, NY 14627, USA}
\affiliation{Department of Physics and Astronomy, University of Utah, Salt Lake City, UT 84112, USA}
\affiliation{Dept. of Physics, Chung-Ang University, Seoul 06974, Korea}
\affiliation{Oskar Klein Centre and Dept. of Physics, Stockholm University, SE-10691 Stockholm, Sweden}
\affiliation{Dept. of Physics and Astronomy, Stony Brook University, Stony Brook, NY 11794-3800, USA}
\affiliation{Dept. of Physics, Sungkyunkwan University, Suwon 16419, Republic of Korea}
\affiliation{Institute of Physics, Academia Sinica, Taipei, 11529, Taiwan}
\affiliation{Dept. of Physics and Astronomy, University of Alabama, Tuscaloosa, AL 35487, USA}
\affiliation{Dept. of Astronomy and Astrophysics, Pennsylvania State University, University Park, PA 16802, USA}
\affiliation{Dept. of Physics, Pennsylvania State University, University Park, PA 16802, USA}
\affiliation{Dept. of Physics and Astronomy, Uppsala University, Box 516, SE-75120 Uppsala, Sweden}
\affiliation{Dept. of Physics, University of Wuppertal, D-42119 Wuppertal, Germany}
\affiliation{Deutsches Elektronen-Synchrotron DESY, Platanenallee 6, D-15738 Zeuthen, Germany}

\author[0000-0001-6141-4205]{R. Abbasi}
\affiliation{Department of Physics, Loyola University Chicago, Chicago, IL 60660, USA}

\author[0000-0001-8952-588X]{M. Ackermann}
\affiliation{Deutsches Elektronen-Synchrotron DESY, Platanenallee 6, D-15738 Zeuthen, Germany}

\author{J. Adams}
\affiliation{Dept. of Physics and Astronomy, University of Canterbury, Private Bag 4800, Christchurch, New Zealand}

\author[0000-0002-9714-8866]{S. K. Agarwalla}
\altaffiliation{also at Institute of Physics, Sachivalaya Marg, Sainik School Post, Bhubaneswar 751005, India}
\affiliation{Dept. of Physics and Wisconsin IceCube Particle Astrophysics Center, University of Wisconsin{\textemdash}Madison, Madison, WI 53706, USA}

\author[0000-0003-2252-9514]{J. A. Aguilar}
\affiliation{Universit{\'e} Libre de Bruxelles, Science Faculty CP230, B-1050 Brussels, Belgium}

\author[0000-0003-0709-5631]{M. Ahlers}
\affiliation{Niels Bohr Institute, University of Copenhagen, DK-2100 Copenhagen, Denmark}

\author[0000-0002-9534-9189]{J.M. Alameddine}
\affiliation{Dept. of Physics, TU Dortmund University, D-44221 Dortmund, Germany}

\author{N. M. Amin}
\affiliation{Bartol Research Institute and Dept. of Physics and Astronomy, University of Delaware, Newark, DE 19716, USA}

\author[0000-0001-9394-0007]{K. Andeen}
\affiliation{Department of Physics, Marquette University, Milwaukee, WI 53201, USA}

\author[0000-0003-2039-4724]{G. Anton}
\affiliation{Erlangen Centre for Astroparticle Physics, Friedrich-Alexander-Universit{\"a}t Erlangen-N{\"u}rnberg, D-91058 Erlangen, Germany}

% ...

\collaboration{407}{IceCube Collaboration}

\begin{abstract}
Neutrino flares in the sky are searched in data collected by IceCube between May 2011 and 2021. 
This data set contains cascade-like events originating from charged-current electron neutrino and tau neutrino interactions and all-flavor neutral-current interactions. 
IceCube's previous all-sky searches for neutrino flares used data sets consisting of track-like events originating from charged-current muon neutrino interactions. 
The cascade data set is statistically independent of the track data sets, and while inferior in angular resolution, the low-background nature makes it competitive and complimentary to previous searches.  
No statistically significant flare of neutrino emission was observed in an all-sky scan. Upper limits are calculated on neutrino flares of varying duration from 1 hour to 100 days. Furthermore, constraints on the contribution of these flares to the diffuse astrophysical neutrino flux are presented, showing that multiple unresolved transient sources may contribute to the diffuse astrophysical neutrino flux. 
\\

\end{abstract}

\section{Introduction} \label{sec:intro}
The astrophysical processes responsible for producing, accelerating, and propagating high-energy cosmic rays have not been resolved. High-energy neutrinos can provide insights into the origins of cosmic rays. In 2013, IceCube reported observations of the diffuse astrophysical neutrino flux \citep{2013}. However, the majority of the diffuse astrophysical neutrino flux still has unresolved origins. Since then, IceCube has observed evidence of time-dependent and steady-state neutrino emission from astrophysical objects. In 2017, IceCube detected a 290 TeV neutrino (IceCube-170922A) in spatial coincidence with the blazar TXS 0506+056; IceCube-170922A was also temporally coincident with enhanced multi-wavelength activity in the blazar \citep{TXS_1}. The statistical significance of this spatial and temporal coincidence was reported at 3$\sigma$. An archival search of 9.5 years of IceCube data found further evidence at a statistical significance of 3.5$\sigma$ for a neutrino flare that occurred between September 2014 and March 2015 that was not coincident with gamma-ray emission \citep{TXS_2}. Since then, further searches~\citep{baikal, ANTARES} have hinted at the possible connection of neutrino flares and blazars. IceCube observed evidence for continuous neutrino emission from the Seyfert galaxy NGC 1068 with a significance of 4.2$\sigma$ \citep{NGC1068}. Recently, IceCube has also observed steady diffuse neutrino emission from the Milky Way at a significance of 4.5$\sigma$ \citep{GP_paper}. In the past decade, time-domain multi-messenger astronomy has observed multiple breakthroughs that serve potential insight into which sources are capable of contributing to the diffuse astrophysical neutrino flux. Not all astrophysical transient events are expected to produce a neutrino flux; however, hadronic astrophysical transient events are primary candidates for contributing to the diffuse neutrino flux \citep{Murase_2019}. In order to remain model-independent, this analysis excludes multi-messenger information and solely uses 10 years of IceCube's cascade data to perform a time-dependent search for neutrino flares across the entire sky. 

The IceCube Neutrino Observatory is located at the geographic South Pole and occupies a cubic kilometer of ice instrumented with 5,160 digital optical modules (DOMs). The DOMs are situated on 86 readout and support cables that have been frozen into the Antarctic glacier at depths between 1450 m -- 2450 m \citep{DAQ, detector}. The primary in-ice DOMs array is composed of 78 readout and support cables; the DOMs have a vertical spacing of 17 m and a horizontal spacing of 125 m. This design allows IceCube to detect the Cherenkov radiation from charged particles that are created by neutrinos with energies from 100 GeV -- 10 PeV interacting with the Antarctic ice. The remaining 8 cables were used to create a higher density sub-volume of IceCube in order to detect neutrinos down to 10 GeV \citep{Deepcore}. Since 2011, IceCube has been fully operational and taking data with $>$ 99\% uptime \citep{detector}.

\section{Search for Neutrino Flares} \label{methods}

\subsection{Cascade events: A new avuenue to search for flares}
IceCube detects electron neutrinos ($\nu_{e}$), muon neutrinos ($\nu_{\mu}$), and tau neutrinos ($\nu_{\tau}$), as well as their antiparticles. At Earth, we expect a flavor ratio ($\nu_{e}$ : $\nu_{\mu}$ : $\nu_{\tau}$) of approximately 1:1:1 due to neutrino oscillations occurring over astronomical distances \citep{Flavor_ratio}. Depending on the interaction type, a neutrino will primarily produce two event topologies: cascades and tracks \citep{energy_reco}. Cascade events are produced by all-flavor neutral-current interactions, as well as charged-current $\nu_{e}$ and $\nu_{\tau}$ interactions, which can produce hadronic and electromagnetic showers~\citep{energy_reco}. Due to the spatial extent of the showers and the scattering length of light within the Antarctic ice, which is shorter compared to the DOM spacing, cascade events appear nearly symmetric and have reconstructed angular resolutions of $\sim$5$^\circ$ -- 15$^\circ$ depending on the energy of the neutrino \citep{Aartsen_2013, GP_paper}. Conversely, tracks are produced when $\nu_{\mu}$ interact with the ice through the charged-current channel and produce a muon that can travel for several kilometers while emitting Cherenkov light. At high energies, tracks have reconstructed angular resolutions $\lesssim$ 1$^\circ$ \citep{10yr_Tracks}. 

However, a major disadvantage of the track data set is the much higher background rates compared to the cascade data set. Atmospheric muons are detected at $\mathcal{O}(10^{9})$  higher rate than signal neutrinos at trigger level, and most of the atmospheric neutrinos that reach the IceCube detector are muon neutrinos. Thus atmospheric backgrounds overwhelmingly produce track events. By using cascade events, the background contamination of the data set is greatly reduced~\citep{GP_paper}. This has a large effect on sensitivities in the Southern Sky, where IceCube does not have the Earth to shield the detector from atmospheric muons. Because the atmospheric background fluxes are expected to be a softer spectrum of $E^{-3.7}$ compared to the neutrino spectrum from astrophysical flares at $E^{-2}$ to $E^{-3}$, the reduction in background effectively lowers the energy threshold of the cascade analysis relative to the track analysis. 

\begin{figure}
    \centering
    \includegraphics[width=0.5\linewidth]{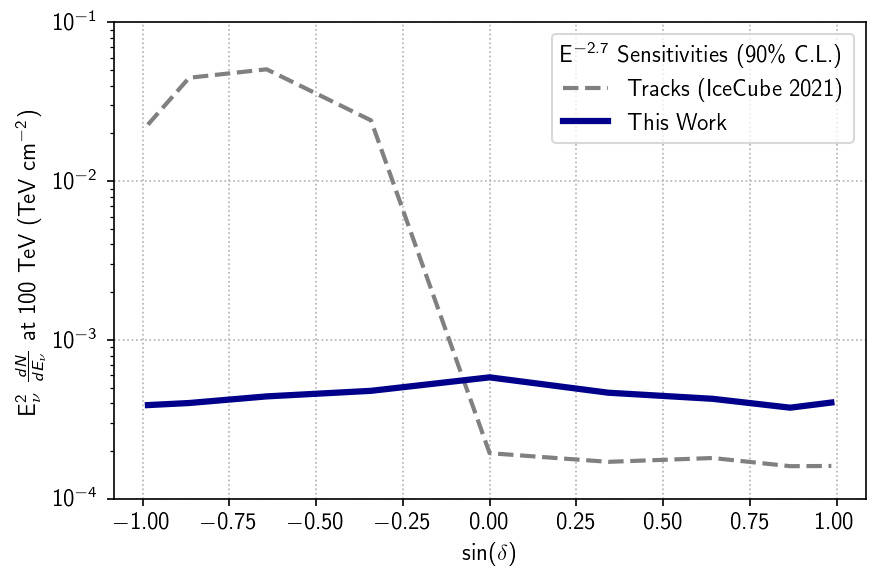}
    \caption{The 90\% C.L. sensitivity flux to an $E^{-2.7}$ neutrino flare emission with a duration of 10 days at various declinations. Using IceCube's track data set~\citep{track_data_release}, the sensitivity deteriorates rapidly in the Southern Sky as the Earth shielding to atmospheric muons disappears, while cascade data set remains flat. Even in the Northern Sky, the sensitivity fluxes remain relatively close because of the soft emission of $E^{-2.7}$ flare that is assumed here. This is due to the low background nature of the cascade data set, which effectively lowers the energy threshold of the analysis as described in the main text. }
    \label{fig:track_comp}
\end{figure}

Fig.~\ref{fig:track_comp} compares the 90\% confidence level (C.L.) sensitivity fluxes of neutrino flares as detected in a track data set and the cascade data set. While track events have better sensitivity in the Northern Sky, the sensitivity deteriorates rapidly in the Southern Sky as Earth shielding of atmospheric muons disappears. The cascade data set remains flat in sensitivity flux since atmospheric muons are not the main background in this data set. Even in the Northern Sky, the sensitivity fluxes remain relatively close when a soft emission of $E^{-2.7}$ flare is assumed, and will become the most sensitive at even softer spectra. 

A region of particular interest to neutrino flare searches is the Galactic Center. The Galactic Center region hosts a supermassive black hole (SMBH) at the position of Sgr A* which is the nearest object capable of strong flare activity. SMBHs can be sources of flare-like emission of cosmic rays and their secondaries, neutrinos and gamma-rays. By using cascade events, we lower the sensitivity flux to such flares by more than an order of magnitude for the emission duration and spectral index assumed in Fig.~\ref{fig:track_comp}. Astrophysical sources of PeV-scale cosmic ray production, dubbed PeVatrons, are hypothesized along the Galactic Plane, with most of the population in the Southern Sky~\citep{Cristofari:2021jkl, app13116433, 2021Natur.594...33C, Bustamante_2023}. Thus the enhancement in neutrino flare sensitivities in the Southern Sky opens a new phase space in Galactic neutrino flare searches. Some models on neutrino emission from from PeVatrons, such as Galactic gamma-ray binaries~\citep{Bykov_2021}, show fluxes in the TeV range and beyond compatible with IceCube's sensitivity ranges.

A previous IceCube search for a neutrino flares yielded no statistically significant observation~\citep{flares_2012_2017}. This analysis also scanned the entire sky for the most statistically significant neutrino flares in the Northern and Southern Skies, and did not utilize external triggers, such as alerts from gamma-ray telescopes. Such an ``untriggered" search can detect astrophysical phenomena that solely produce neutrinos or produce neutrinos and other astrophysical messengers at times offset from the detection of optical, X-ray, or gamma-ray flares. In IceCube, when searching for neutrino flares that have durations of \begin{math}\mathcal{O}(10^{2})\end{math}days or less, this method becomes more sensitive than time-integrated searches \citep{Braun_2010}. IceCube's previous untriggered analyses only used track events due to their high event rate and small angular uncertainties. The analysis of cascade events provides a new opportunity to use a statistically independent data set to observe the transient sky. This is the first such all-sky search utilizing a cascade data set. 

\subsection{Cascade Data set}
The cascade data set contains 59,592 cascade-like events detected with IceCube's complete 86-string configuration from May 13, 2011 to May 27, 2021 with energies between 500 GeV -- 5.35 PeV. The angular resolution of the data set is energy-dependent. At 1 TeV, the median angular resolution is $>$ 15$^\circ$; at 100 TeV, the median angular resolution is $\lesssim$ 10$^\circ$ and improves to $\sim$ 5$^\circ$ at PeV energies. The data set was designed to search for time-integrated all-sky and Galactic diffuse emission~\citep{GP_paper}, and uses novel event reconstruction and event selection techniques with neural networks and boosted decision trees~\citep{dnn_reco}. Approximately 87\% of the cascade events are estimated to be atmospheric neutrinos, 7\% are astrophysical neutrinos, and the remainder being atmospheric muons \citep{GP_paper}. The atmospheric neutrino spectrum is comparatively softer than that of astrophysical neutrinos \citep{Aartsen_2015}. Thus, at energies between 10 TeV -- 100 TeV, atmospheric neutrinos become the subdominant component of the cascade data set \citep{GP_paper}.

\subsection{Analysis Methods}
\label{subsection:analysismethods}

We use 10 years of cascade events to search for the most statistically significant spatial and temporal clustering of events in IceCube's Northern ($\delta$ $>$ -5$^\circ$) and Southern ($\delta$ $<$ -5$^\circ$) Sky, which were predefined in a blind analysis. The north/south split was chosen to be consistent with previous IceCube analyses using track events. The points within 10$^\circ$ of the celestial poles are excluded due to their low statistics and limited background estimation. The clustering search uses a maximum likelihood method to identify time-dependent neutrino emission from point sources, similar to IceCube's previous analyses \citep{Braun_2010, flares_2012_2017}. The likelihood,

\begin{equation} \label{eqn:likelihood}
    \mathcal{L} =  \prod_i^N ( \frac{n_s}{N} S_i + ( 1 - \frac{n_s}{N} ) B_i ),
\end{equation}

\noindent runs over every $i$th event in the data set which includes $N$ events. Here, $S_i$ and $B_i$ are probability density functions (PDFs) which are products of individual spatial, energy, and temporal PDFs. To search for neutrino emission, the sky is partitioned into 12,288 pixels of equal solid angle using HEALPix\footnote{https://healpix.jpl.nasa.gov/} \citep{healpix}. The spatial PDF accounts for the angular distance between the center of each pixel and the reconstructed arrival direction of each cascade event. Due to the rotation of the Earth, the angular distribution of background events is assumed to be uniform in right ascension. We also assume a Gaussian PDF to test for temporal clustering between given cascade events and and treat IceCube’s background event rate as uniform in time while also accounting for detector livetime. IceCube's previous searches have been performed with both a Gaussian and a ``box-shaped'' temporal PDF which follows a Heaviside step function. Here we use the Gaussian PDF since it is less computationally intensive. The energy PDFs account for the astrophysical spectra and the energy distribution of the cascade events as a function of declination. We assume that for the signal hypothesis, the neutrino flux follows an unbroken power-law, E$^{-\gamma}$, where $\gamma$ is the spectral index.

In this analysis, we define the null hypothesis such that there is no spatial and temporal neutrino clustering for a given area of the sky. The alternative hypothesis fits for the following four parameters at every pixel: number of signal events ($n_{s}$), which is constrained to be $n_s\geq0$, signal spectral index ($\gamma$), allowed to fit within the range $1.0\leq\gamma\leq4.0$, mean time of the flare ($T_{0}$ [Modified Julian Date (MJD)]), allowed to fit to any time during the livetime of the data set, and half-width flare duration ($\sigma_{t}$ [days]), constrained to $\sigma_{t}\geq10^{-11}$ and maximally to half of the duration of the livetime of the data set, in order to search for neutrinos that are spatially and temporally clustered. The alternative hypothesis indicates an excess of signal-like events that surpass the expected background. To test the null and alternate hypothesis we use the likelihood ratio test, as described in \cite{Wilks}, to construct a test statistic (TS),

\begin{equation} \label{ts}
    \textrm{TS} = -2 \log  \left[ \frac{T_{\mathrm{max}} - T_{\mathrm{min}}}{\sigma_t} \times \frac{\mathcal{L}(n_s=0)}{\mathcal{L}(n_s, \gamma, T_0, \sigma_t)} \right]. 
\end{equation}

\noindent In \autoref{ts}, $T_\mathrm{max}$ and $T_\mathrm{min}$ refer to the livetime bounds of the cascade data set. The TS includes a marginalization term for the flare duration in addition to the likelihood ratio; this prevents a bias that arises from over-fitting for $\sigma_{t}$ with shorter durations. IceCube's previous untriggered flare searches have searched for flares by imposing a signal-over-background ($S/B$) threshold to reduce the computational complexity of the analysis. The $S/B$ threshold is used to seed potential flares with events that have a high probability of being a signal, based off their reconstructed direction and energy. Thus, $S/B$ only accounts for the spatial and energy PDFs of a given event. Previous all-sky track analyses had data sets with high data rates; these analyses used thresholds of \begin{math}S/B > 1000\end{math}to efficiently search for flares. Since the cascade data set has an order of magnitude fewer events than IceCube's track data set, we relax this threshold to \begin{math}S/B > 1\end{math}. 

To calculate the TS, the events' arrival directions are uniformly randomized in right ascension; the time of the event is also randomized. A TS is then calculated for each pixel in the all-sky search. This process is repeated \begin{math}\mathcal{O}(10^{4})\end{math} -- \begin{math}\mathcal{O}(10^{5})\end{math}times in order to obtain an ensemble of TS that follows the null hypothesis of no spatial or temporal clustering. Each TS in the ensemble represents a different realization of background data. To minimize the computational expense of building an ensemble of TS for each pixel, and to account for the declination dependence of the data, we build an ensemble of TS in 81 declination bins and one right ascension bin. The declination bins range from [-80$^\circ$, 80$^\circ$] and are evenly spaced in 2$^\circ$ increments. The TS for the data, TS$_\mathrm{data}$, is then calculated for every pixel and compared to the corresponding ensemble of TS in order to calculate a local p-value ($p_\mathrm{local}$), which is the probability of getting a TS $\geq$ TS$_\mathrm{data}$ from a random background realization of our data. We report two local p-values, $p_\mathrm{local}$, for the data that correspond to the most statistically significant pixels in the Northern and Southern Sky, respectively. Since we test many pixels on the sky, the two local p-values, $p_\mathrm{local}$, for the Northern and Southern skies are corrected for statistical trials to report global p-values ($p_\mathrm{global}$). We construct $p_\mathrm{global}$ by considering \begin{math}\mathcal{O}(10^{2})\end{math}randomized sky scans that correspond to different realizations of background data, and then taking the most statistically significant pixels in the Northern and Southern Sky from each sky scan in order to obtain respective distributions of the most statistically significant $p_\mathrm{local}$. Finally, the two values of $p_\mathrm{global}$ are reported as the trial-corrected significances.

\section{Results \& Implications} \label{results}
No significant spatial and temporal clustering of astrophysical neutrinos was observed in this analysis. 


In the Northern Sky, the most significant point is located at \begin{math} (\alpha, \delta) = (322.03^{\circ}, -4.78^{\circ}) \end{math}with a corresponding $p_\mathrm{local}$\begin{math}=2.45\times 10^{-3} \end{math}and a pre-trials significance of \begin{math}2.81\sigma.\end{math}The most significant point in the Southern Sky is located at \begin{math} (\alpha, \delta) = (320.63^{\circ}, -5.98^{\circ}) \end{math}with a respective $p_\mathrm{local}$\begin{math}=2.12\times 10^{-3} \end{math}and a pre-trials significance of \begin{math}2.86\sigma. \end{math}The ``hottest'' spots are separated by $<2^\circ$ in declination and are located in the same extended ``warm'' spot. This is due in part to the angular resolution of the cascade events. Since cascade events are correlated with different pixels, the statistical fluctuations near the hottest spot have similar $p_\mathrm{local}$. After correcting trials, we report the final p-values of $p_\mathrm{global} =$ 0.71 in the Northern Sky, and $p_\mathrm{global} =$ 0.51 in the Southern Sky. The results of this all-sky search for transient astrophysical neutrino emission are consistent with the background-only hypothesis.

\subsection{Diffuse Flux Constraints}

\begin{figure}[h!]
    \centering
        \includegraphics[width=.49\linewidth]{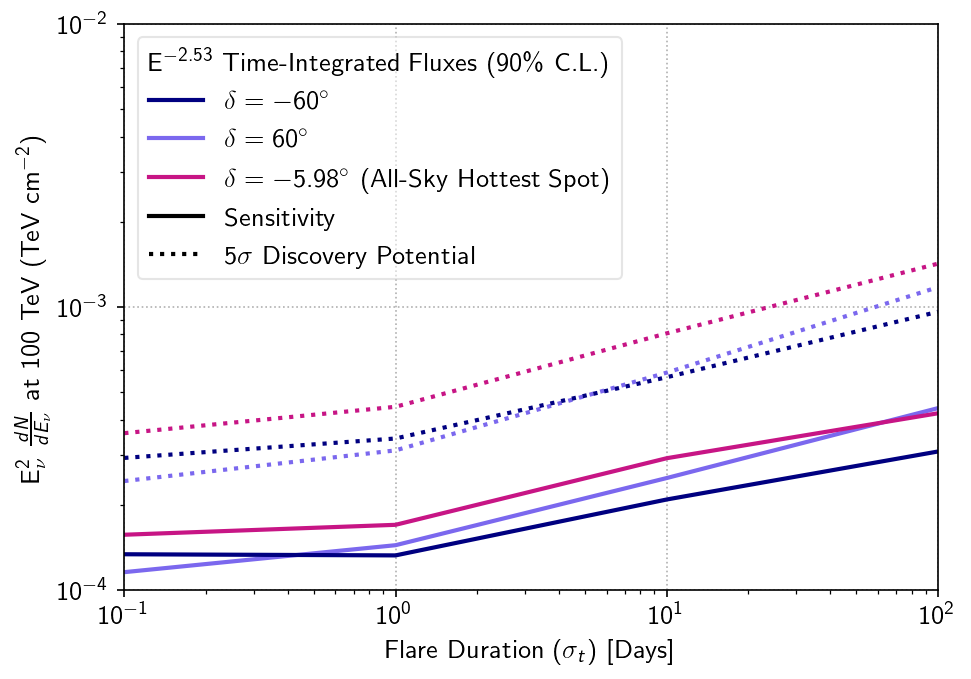}
    \includegraphics[width=.45\linewidth]{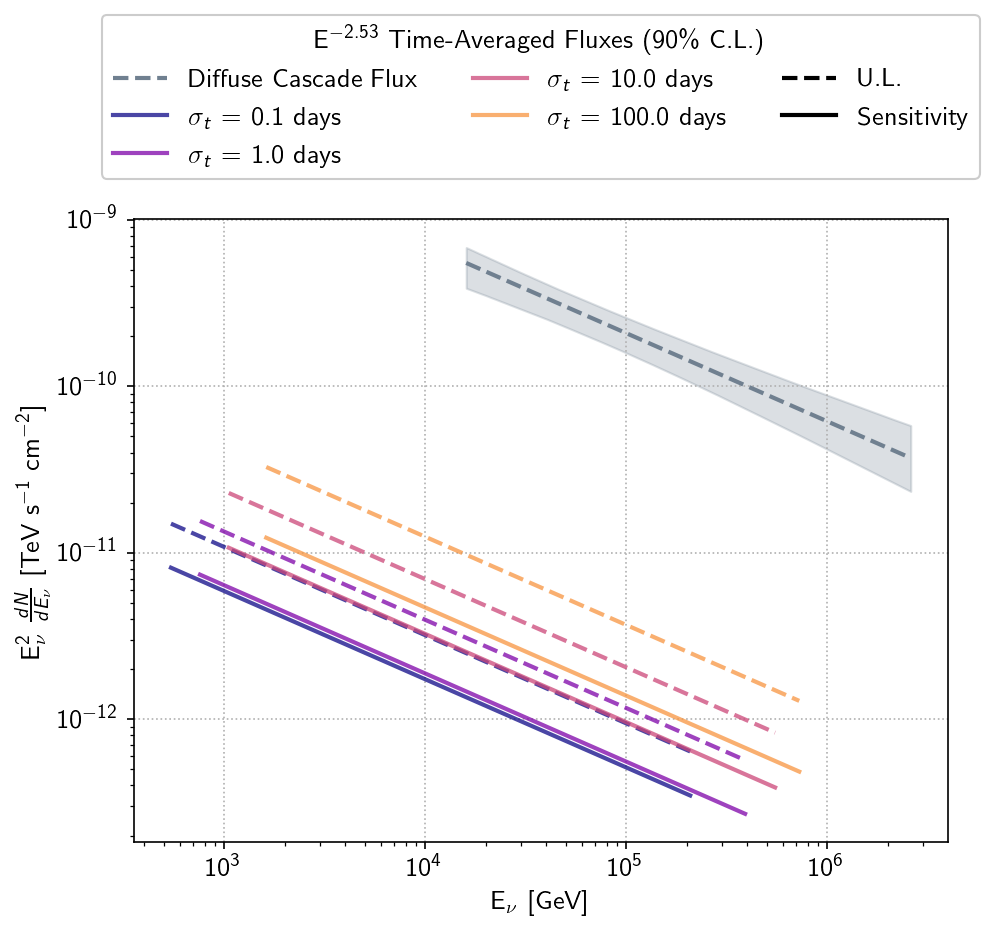}
    \caption{\textbf{Left: } Fluences of neutrino flares at the 90\% sensitivity and 5$\sigma$ discovery potential of this analysis, assuming flare durations between 0.1 days and 100 days. An energy spectrum of $E^{-2.53}$ is assumed in these flares to match that of the measured diffuse neutrino emission in cascades \citep{diffuse_cascade}. Simulated flares are placed at the declination of the hottest spot found, and also at a benchmark declination of $\pm60^{o}$. \textbf{Right:} Comparison of the time-averaged upper limit and sensitivity fluxes, at the declination of the hottest spot, to the previously measured steady-state diffuse neutrino flux~\citep{diffuse_cascade}. As the flare duration increases, the maximum possible contribution to the diffuse cascade flux increases. For each possible flare duration shown, hundreds or thousands of independent neutrino flares are needed to equal the entire diffuse astrophysical flux. At the longest flare duration (100 days), at least 100 neutrino flares are needed to account for the observed diffuse flux.}
    \label{fig:diffuse_limits}
\end{figure}

The sensitivity and discovery potential fluences at the hottest spot, along with two benchmark declinations of $\pm60^{o}$, are shown as a function of $\sigma_{t}$ in \autoref{fig:diffuse_limits} -- left. IceCube has previously measured a time-integrated diffuse astrophysical neutrino flux with cascade events \citep{diffuse_cascade}. This plot assumes the flare emission has a power law of E$^{-2.53}$ taken from the best fit spectral index of the measured diffuse cascade flux. 

As no significant flare was observed, we calculated the 90\% upper limits (U.L.) on neutrino fluxes (\autoref{fig:diffuse_limits} -- right) for the all-sky hottest spot and we discuss implications for the observed diffuse astrophysical flux. Assuming a power-law of E$^{-2.53}$ taken from the measured diffuse cascade flux, we calculate the time-averaged fluxes at the sensitivity of this analysis and the resulting U.L. from data, each at the 90\% C.L., at the location of the most significant point. The energy range used for each flux corresponds to the energy range of the neutrino events contributing to 90\% of the flux. Fluxes are converted to time-averaged fluxes by distributing them over the entire livetime of the data set in order to compare to the measured diffuse cascade flux. As the flare duration increases, the maximum possible contribution from a single neutrino flare to the diffuse astrophysical neutrino flux increases. However, at least 100 neutrino flares of $\sigma_{t}$ = 100 days are needed to account for the diffuse astrophysical flux. 

\section{conclusions}
In the all-sky search for transient neutrino emission, no significant spatial and temporal clustering was observed. We provide upper limits on the time-averaged neutrino flux at the 90\% confidence level for various assumed flare durations, as shown in \autoref{fig:diffuse_limits} -- left. These all-sky upper limits show that depending on the flare duration hundreds or thousands of individual neutrino flares are needed in order to comprise the diffuse astrophysical neutrino flux. These all-sky upper limits are not dependent on a specified class of astrophysical transient objects; thus, multiple unresolved transient objects may contribute to the diffuse neutrino flux.

Cascades provide upper limits that can help constrain future searches for transient neutrino emission. Cascade data sets are considered to be independent of IceCube's track data sets, and future searches for transient neutrino emission could combine tracks and cascades. In addition, the next generation of IceCube, IceCube-Gen2, will have the capability to conduct a more sensitive search for transient neutrino emission across the entire sky \citep{gen2}. This will provide us with the opportunity to observe the origin of high-energy astrophysical neutrinos and further motivate time-domain multi-messenger astronomy.

\section*{Acknowledgements}
The IceCube collaboration acknowledges the significant contributions to this manuscript from Michael Kovacevich.
USA {\textendash} U.S. National Science Foundation-Office of Polar Programs,
U.S. National Science Foundation-Physics Division,
U.S. National Science Foundation-EPSCoR,
U.S. National Science Foundation-Office of Advanced Cyberinfrastructure,
Wisconsin Alumni Research Foundation,
Center for High Throughput Computing (CHTC) at the University of Wisconsin{\textendash}Madison,
Open Science Grid (OSG),
Partnership to Advance Throughput Computing (PATh),
Advanced Cyberinfrastructure Coordination Ecosystem: Services {\&} Support (ACCESS),
Frontera computing project at the Texas Advanced Computing Center,
U.S. Department of Energy-National Energy Research Scientific Computing Center,
Particle astrophysics research computing center at the University of Maryland,
Institute for Cyber-Enabled Research at Michigan State University,
Astroparticle physics computational facility at Marquette University,
NVIDIA Corporation,
and Google Cloud Platform;
Belgium {\textendash} Funds for Scientific Research (FRS-FNRS and FWO),
FWO Odysseus and Big Science programmes,
and Belgian Federal Science Policy Office (Belspo);
Germany {\textendash} Bundesministerium f{\"u}r Bildung und Forschung (BMBF),
Deutsche Forschungsgemeinschaft (DFG),
Helmholtz Alliance for Astroparticle Physics (HAP),
Initiative and Networking Fund of the Helmholtz Association,
Deutsches Elektronen Synchrotron (DESY),
and High Performance Computing cluster of the RWTH Aachen;
Sweden {\textendash} Swedish Research Council,
Swedish Polar Research Secretariat,
Swedish National Infrastructure for Computing (SNIC),
and Knut and Alice Wallenberg Foundation;
European Union {\textendash} EGI Advanced Computing for research;
Australia {\textendash} Australian Research Council;
Canada {\textendash} Natural Sciences and Engineering Research Council of Canada,
Calcul Qu{\'e}bec, Compute Ontario, Canada Foundation for Innovation, WestGrid, and Digital Research Alliance of Canada;
Denmark {\textendash} Villum Fonden, Carlsberg Foundation, and European Commission;
New Zealand {\textendash} Marsden Fund;
Japan {\textendash} Japan Society for Promotion of Science (JSPS)
and Institute for Global Prominent Research (IGPR) of Chiba University;
Korea {\textendash} National Research Foundation of Korea (NRF);
Switzerland {\textendash} Swiss National Science Foundation (SNSF).

\bibliography{v6.2}
\bibliographystyle{aasjournal}

\end{document}